**Ridge-filter crosstalk in conformal proton FLASH planning: dependence on beamlet pitch and iterative mitigation**

Zongsheng Hu, Ph.D.[1]; Yuting Li, Ph.D.[1]; Henry Meyer, B. A.[1]; Xiaochun Wang, Ph.D.[1]; Susan L. McGovern, M.D., Ph.D.[2]; Emil Schüler, Ph.D.[1]; Radhe Mohan, Ph.D.[1]; Uwe Titt, Ph.D.[1]

1. Department of Radiation Physics, The University of Texas MD Anderson Cancer Center, Houston, Texas
2. Department of Radiation Oncology, The University of Texas MD Anderson Cancer Center, Houston, Texas

**Corresponding Author:** Zongsheng Hu

**Conflict of Interest Statement:** None

**Acknowledgement Statement:** This work is supported in part by Cancer Center Support Grant P30 CA016672 from the NCI of the NIH to The UT MDACC. We acknowledge the support of the High Performance Computing for research facility at The University of Texas MD Anderson Cancer Center for providing computational resources that have contributed to the research results reported in this paper. We thank Ashli Nguyen-Villarreal, Associate Scientific Editor, and Joe Munch, Senior Scientific Editor, in the Research Medical Library at The University of Texas MD Anderson Cancer Center for editing this article.

**Author Contributions**

Zongsheng Hu, Ph.D. designed the computational framework, implemented all simulation and optimization codes, performed data analysis, and drafted the manuscript.

Uwe Titt, Ph.D., Radhe Mohan, Ph.D., Emil Schüler, Ph.D., Susan L. McGovern, M.D., Ph.D., and Zongsheng Hu, M.S. jointly conceived the study and shaped the overall research direction.

Susan L. McGovern, M.D., Ph.D. provided the clinical patient case, treatment prescription, and clinical interpretation relevant to the study.

Yuting Li, Ph.D., and Xiaochun Wang, Ph.D. provided guidance on treatment planning methodology, clinical workflow considerations, and interpretation of dosimetric results.

Henry Meyer, B.A. assisted with data processing, figure generation, and contributed to manuscript editing.

**ABSTRACT**

**Objective:**

Patient-specific ridge filters (PSRFs) have been proposed to enable conformal single-energy proton FLASH dose delivery without energy switching. However, converting an optimized spot-based dose distribution into physically adjacent ridge-filter structures can introduce inter-beamlet modulation errors that are not captured by conventional isolated-spot optimization. In this work, we investigated ridge-filter (RF) crosstalk as a source of these discrepancies, characterized its dependence on the beam-width-to-pitch relationship, and developed an iterative mitigation strategy.

**Approach:**

A dose influence matrix was calculated using Monte Carlo simulation for monoenergetic proton beamlets passing through RFs with different thicknesses. A baseline spot-weighted intensity-modulated proton therapy (IMPT) plan was generated to satisfy the dose constraints, and the optimized spot weights were converted into PSRF geometries. The corresponding PSRF dose distributions were then calculated by explicitly modeling the PSRF in the beam path of the scanned beamlets. RF crosstalk was quantified by comparing PSRF and baseline IMPT plans. To assess the dependence of crosstalk on beamlet spacing, lateral spacings of 8, 10, 12, and 15 mm were evaluated. Dose agreement was assessed using gamma analysis, dose-volume histogram metrics, and mean relative dose difference. An iterative re-optimization method was then applied to compensate for dose discrepancies caused by RF crosstalk and was further evaluated in both water-phantom and patient CT geometries.

**Results:**

RF crosstalk introduced hot and cold spots that degraded agreement between the PSRF and baseline IMPT plans. For the same beam spot size and target geometry, the magnitude of this effect increased as beamlet spacing decreased. Iterative re-optimization substantially mitigated RF crosstalk, reducing the mean relative dose difference within the target from 8.9% to 3.4% in the water phantom and from 3.7% to 1.8% in the CT geometry.

**Significance:**

RF crosstalk is an important and previously under-characterized source of dose inconsistency in ridge-filter-based conformal proton FLASH planning. These results show that the effect depends strongly on the beam-width-to-pitch relationship and can be mitigated through iterative re-optimization, providing a practical framework for improving the dosimetric accuracy and robustness of patient-specific single-energy proton FLASH dose delivery.

## INTRODUCTION

Ultra-high dose rate, or FLASH radiotherapy (RT), typically defined as delivery at dose rates exceeding 40 Gy $s^{-1}$, is an emerging modality that has shown remarkable normal tissue sparing with tumor control equivalent to that of conventional dose rate RT in preclinical models, motivating substantial interest in its clinical translation[1-8]. Proton beams are particularly attractive for FLASH radiotherapy because of their finite range and Bragg peak characteristics, which enable the delivery of dose to deep-seated targets with improved normal tissue dose sparing compared with electron and photon beams[9-22]. Thus, proton FLASH (pFLASH) RT is an appealing approach for deep-seated complex target shapes, which are frequent and clinically relevant.

Current pFLASH RT–capable facilities are mostly limited to small animal experiments due to technological limitations [23]. To advance the clinical translation of pFLASH RT, it is imperative to gather data from mid-sized and large animals and from clinical trials. Therefore, the development of a clinically practical pFLASH RT system capable of providing conformal and large target volume coverage is crucial.

One proposed approach to achieving conformal pFLASH dose delivery is the use of single-energy pencil beam scanning combined with patient-specific ridge filter (PSRF) devices. In these approaches, depth-dose modulation is achieved locally for each beamlet with beamlet-specific ridge filters (BSRF) and beamlet-specific range compensator (BSRC), allowing conformal coverage without the need for energy switching that is time consuming and incompatible with maintaining ultra-high dose rates. Several studies have demonstrated the feasibility of such strategies or equivalent concepts, as well as optimization frameworks that account for dose, dose rate, and other LETs[24-26].

Currently, the common approach to creating such PSRF treatment plan is to first generate a baseline intensity-modulated proton therapy (IMPT) plan. The PSRF is then designed by mapping the optimized spot weights from the IMPT plan into the corresponding geometric features of the filter so that the transverse area of each RF layer is proportional to the assigned spot weight [24-26]. However, in IMPT optimization, dose optimization is performed assuming that each pencil beam spot contributes independently to the total dose distribution. When translating an optimized IMPT plan into a PSRF implementation, this assumption is no longer strictly valid. In practice, the PSRF consists of a physically contiguous array of beamlet-specific modulating structures, and the lateral fluence of each pencil beamlet has a gaussian profile, which cannot be fully covered by the corresponding BSRF. As a result, portions of a given beamlet can traverse neighboring BSRF structures, leading to unintended modulation by adjacent beamlet-specific ridge elements. This effect introduces discrepancies between the intended (spot-based IMPT plan) and delivered (device-based PSRF plan) dose distributions.

In this work, we refer to this phenomenon as ridge-filter (RF) crosstalk, defined as the unintended modulation of a beamlet by neighboring ridge-filter structures due to lateral beam spread. Although similar dose inconsistencies have been qualitatively observed in prior ridge-filter–based planning studies, the underlying mechanism has not been systematically isolated or quantitatively characterized. In particular, it remains unclear under what conditions RF crosstalk becomes significant, how it depends on delivery parameters such as beamlet spacing, and to what extent it can be mitigated within a practical planning framework.

A key factor governing RF crosstalk is the relationship between the effective lateral beam width and the beamlet pitch (i.e., lateral spacing of adjacent beamlets). As beamlets are placed closer together to improve dose conformality, increased overlap of their lateral fluence profiles can lead to stronger interaction across adjacent ridge-filter structures, amplifying crosstalk effects. Conversely, increasing beamlet spacing reduces such overlap but may degrade target conformality. This trade-off suggests that RF crosstalk represents an inherent physical limitation in RF-based conformal proton planning, particularly in the context of single-energy FLASH delivery strategies that rely on tightly spaced beamlets.

Despite its potential impact, RF crosstalk has not been systematically analyzed in the context of conformal proton FLASH planning. Moreover, existing approaches that convert optimized spot weights into ridge-filter geometries typically do not explicitly account for these inter-beamlet interactions, which may result in clinically relevant dose discrepancies such as hot and cold spots within the target and OAR. Addressing this limitation is essential to ensure the dosimetric fidelity and robustness of PSRF-based treatment strategies.

The purpose of this study is therefore threefold. First, we identify and quantify RF crosstalk as a source of dose inconsistency in ridge-filter-based conformal proton FLASH planning by comparing dose distributions obtained from spot-based IMPT optimization and corresponding PSRF implementations. Second, we systematically investigate the dependence of RF crosstalk on beamlet lateral spacing under same beam conditions, using a set of representative phantom geometries. Third, we develop and evaluate an iterative re-optimization strategy that compensates for crosstalk-induced discrepancies and restores agreement between the PSRF and baseline dose distributions. The proposed framework is demonstrated in both water-phantom and patient CT–based geometries.

By explicitly characterizing RF crosstalk and providing a practical mitigation approach, this work aims to improve the physical accuracy and robustness of ridge-filter-based conformal proton FLASH planning. More broadly, it highlights the importance of accounting for inter-beamlet interactions when translating idealized spot-based optimization results into physically realizable modulation devices.

## METHODS

### Mini-SOBP scanning strategy and PSRF concept

The overall concept of the single-energy conformal delivery strategy is illustrated in Figure 1. In this strategy, a scanning magnet steers the monoenergetic pencil beam to transversally cover the planning target volume. PSRFs and, when applicable, PSRCs provide depth modulation. The PSRF and PSRC are composites of two-dimensional array of adjacent BSRFs and BSRCs, each corresponding to a single scanning beam position (i.e. a beamlet). Each beamlet-specific ridge filter (BSRF) generates a mini spread-out Bragg peak (mSOBP), and the superposition of all modulated beamlets produces the desired three-dimensional dose distribution. Each BSRF consists of a stack of discrete layers, forming a stepped pyramidal geometry that modulates the longitudinal dose profile. The distal range is adjusted using either beamlet-specific ridge-filter design alone or in combination with a range compensator, depending on the configuration. This design enables conformal dose delivery using a monoenergetic beam without energy switching. A 2-mm layer thickness was selected as a balance between efficiency and modulation resolution: thicker layers improve the efficiency of 3D printing and simplify optimization, whereas thinner layers enhance the ability to modulate the depth-dose distribution but increase 3D printing and computational complexity. Dose distribution along the depth direction is modulated by optimizing the shape of each BSRF. The 3D-printed PSRF and PSRC are composed of a resin with a measured density of 1.22 g $cm^{-3}$ and a water equivalent thickness of 1.185. The PSRF and PSRC could be integrated into a single unit to facilitate efficient 3D printing.

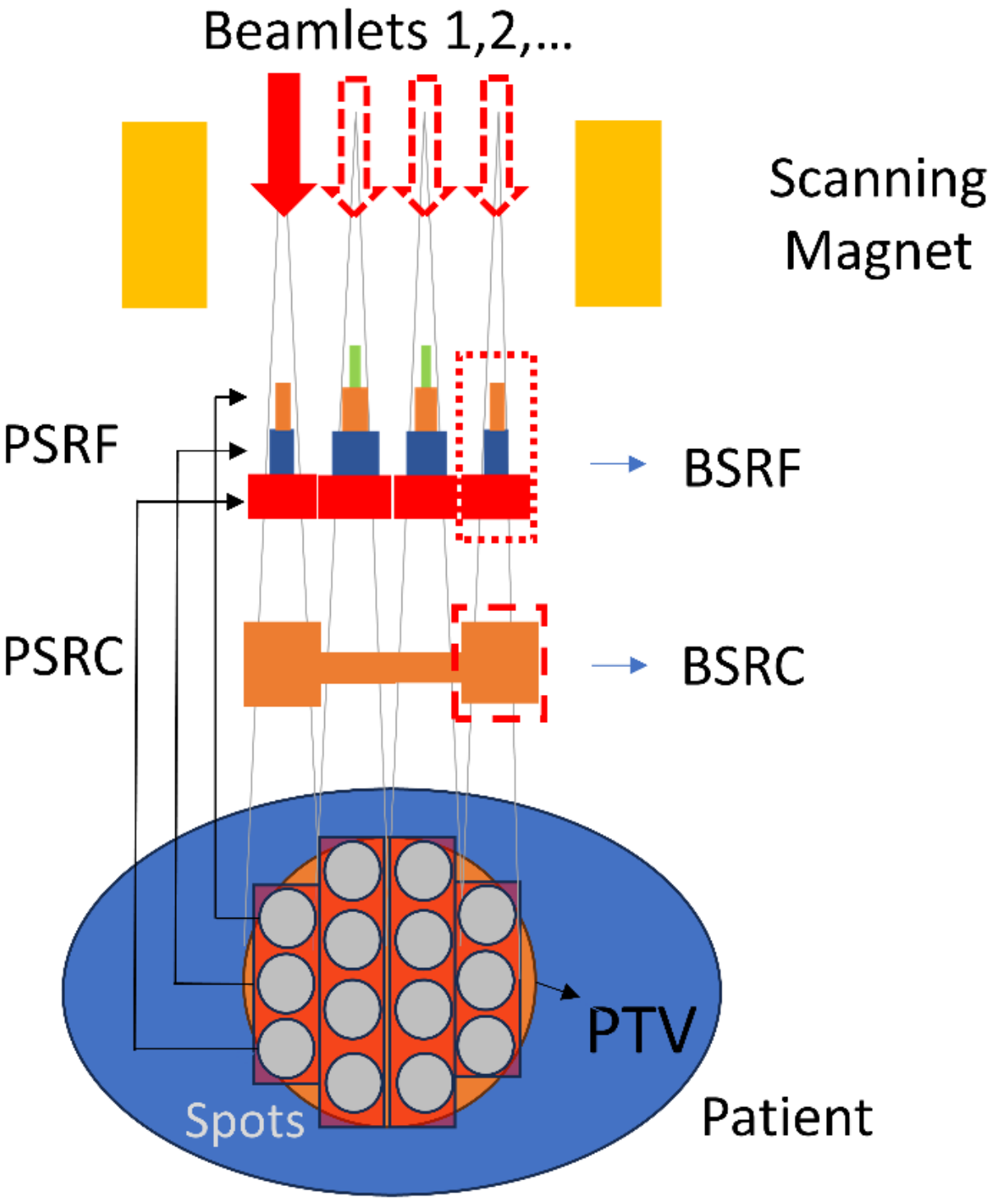

*Figure 1. Schematic design of the mini-spread-out Bragg peak (mSOBP) scanning strategy. A scanning magnet laterally steers monoenergetic pencil beamlets across the target volume. Each beamlet passes through a patient-specific ridge filter (PSRF) and patient-specific range compensator (PSRC), which are composed of beamlet-specific subunits. The beamlet-specific ridge filter (BSRF) modulates the depth-dose distribution, and the beamlet-specific range compensator (BSRC) adjusts the distal range of the Bragg peak. Together, these devices generate mSOBPs that conform to the target. The superposition of many such modulated beamlets creates a desired dose distribution across the planned target volume (PTV). For visualization purposes, only a single pyramidal element is depicted for each beamlet-specific ridge filter (BSRF); in the actual design, each BSRF comprises multiple repetitive pyramids arranged transversely (as shown in Figure 2).*

**Beam model and Monte Carlo simulation**

A monoenergetic proton beam with a nominal energy of 159.5 MeV was used throughout this study. This energy was selected to reach the appropriate distal edge of the target (17.1 cm in water) and match the maximum beam current of the system[27]. The incident beam was modeled with a Gaussian lateral spatial distribution and an energy spectrum calibrated to measured data. The full width at half maximum (FWHM) of the lateral beam profile at the Bragg peak in water was 16.2 mm and 16.7 mm in the x and y directions, respectively. In air, where RF was located were 14.7 mm in x direction and 15.9 mm in y direction. The beam model was validated by comparing MC-simulated depth-dose curves and lateral profiles with measurements.

Monte Carlo simulations were performed using TOPAS with the default proton physics list[28]. Dose was scored on a three-dimensional grid with a voxel size of 1 ×1 × 1 $mm^3$. For each simulation, $10^9$ primary particles were simulated. Water phantoms and patient CT geometries were modeled as voxelized media with appropriate material assignments. Ridge-filter geometries were imported as STL file with a spatial resolution of 0.01 mm.

**Dose influence matrix and IMPT optimization**

To optimize the shape of the PSRF, we calculated an influence matrix using Monte Carlo (MC) simulations, which was derived by computing the 3D dose distributions of monoenergetic pencil beams passing through RF resin materials of different thicknesses. Each column of the influence matrix corresponds to the dose contribution of a specific RF thickness and beam lateral position. Based on this matrix, a baseline IMPT plan was generated by optimizing individual spot weights $w$ using a quasi-Newton method to satisfy typical clinical dose constraints. The dose distribution can be represented schematically as:

$$D = \hat{A} \cdot \vec{w},$$

where $D$ is the resulting 3D dose distribution, $\hat{A}$ is the influence matrix, and $\vec{w}$ is the vector of spot weights derived from the IMPT plan. Optimization was performed by minimizing a cost function consisting of target coverage and organ-at-risk (OAR) sparing terms.

$$min \ \sum_i f_i\big(D(w)\big), \qquad w \geq 0$$

Where $f_i$ represents the constraint functions of each individual target and OAR, $f_i$ can be in the form of maximum/minimum dose, DVHs, dose uniformity, and mean dose constraint function

**Generation and calculation of PSRF plans**

The initial PSRF geometry was then derived from the optimized IMPT spot weights by mapping each weight to the corresponding BSRF structure. Specifically, the transverse area of each ridge layer was defined to be proportional to the magnitude of the spot weight, ensuring that the nominal depth-dose modulation matched the IMPT solution under the assumption of isolated beamlets. The resulting PSRF-based dose distribution was calculated using full MC simulation with the PSRF explicitly included in the beam path during MC calculation.

**Ridge-filter crosstalk: Mechanisms and physical origin**

The origin and mechanism of RF crosstalk effect is illustrated in Figure 2. In the IMPT optimization framework described above, each pencil beam spot is assumed to be modulated only by its corresponding BSRF. However, because each pencil beam's lateral profile is Gaussian-shaped, its contributions extend into neighboring regions. As a result, the lateral parts of the beamlet pass through and are unintentionally modulated by adjacent BSRF sections. This overlap, which we refer to as RF crosstalk, introduces discrepancies between the baseline IMPT plan (which is based on isolated spot weights) and the PSRF plan (which involves physically adjacent BSRF structures). The effect manifests as unwanted hot and cold spots in the dose distribution, reducing homogeneity and conformity. Such inconsistencies have also been observed and reported in other PSRF studies [24-26], highlighting the importance of explicitly modeling and correcting for RF crosstalk.

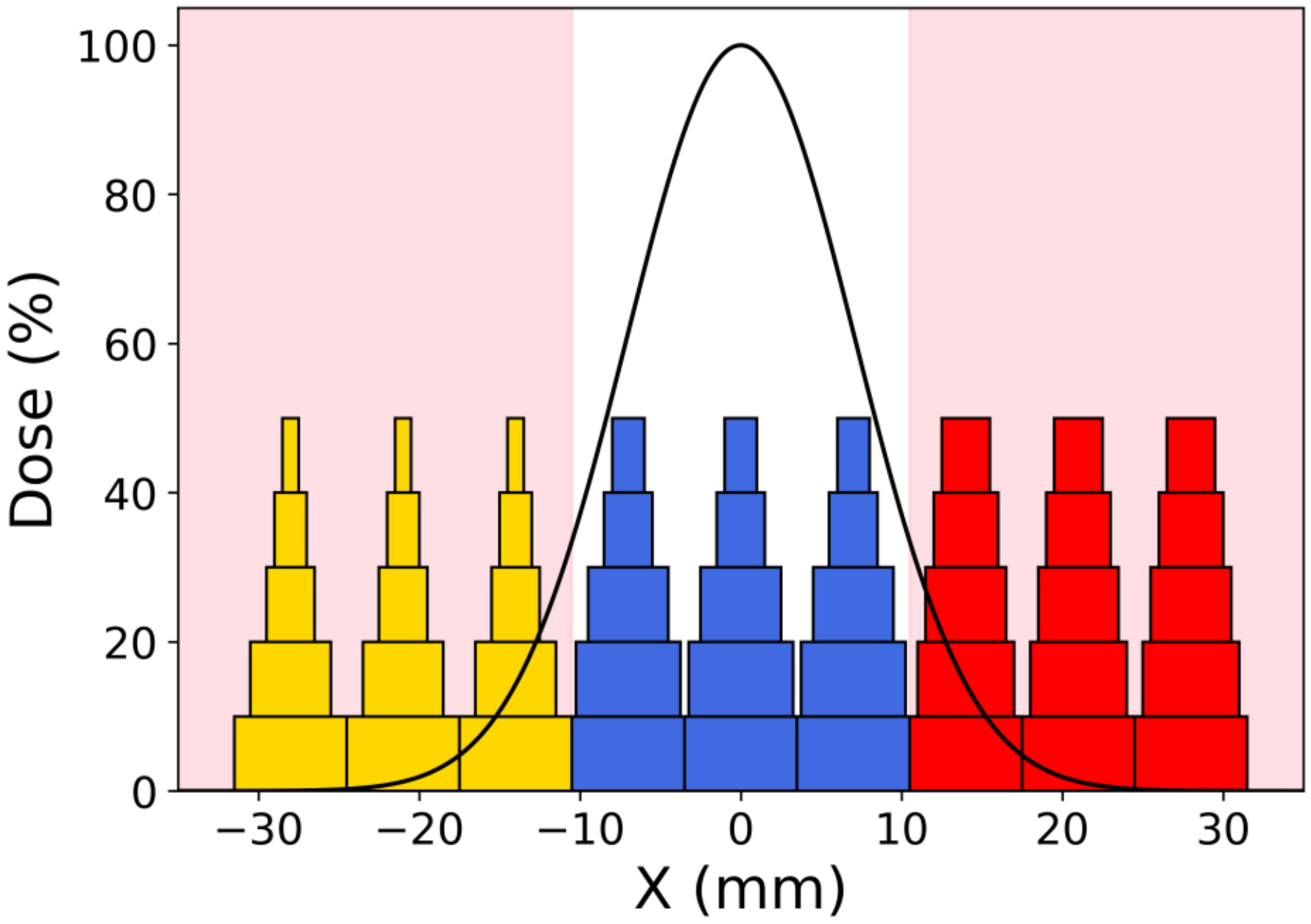

*Figure 2. Illustration of the ridge filter crosstalk effect. The black curve shows the Gaussian lateral profile of a pencil beam. Colored bars represent beamlet-specific ridge filter sections (yellow, blue, red). Although each filter section is designed to shape its corresponding beam spot, the extended tails of the Gaussian profile overlap with adjacent filter sections (pink shading), leading to ridge filter crosstalk and dose discrepancies.*

**Quantification of ridge-filter crosstalk**

RF crosstalk was quantified by comparing the PSRF dose distribution with the corresponding baseline IMPT dose distribution. The mean relative dose error (MRE) within the planning target volume (PTV) was calculated as the average voxel-wise absolute difference normalized to the IMPT dose. In addition, dose–volume histogram (DVH) metrics were used to assess target coverage and dose homogeneity, including D95, D5, D2, and maximum dose. To further evaluate spatial dose agreement, a three-dimensional gamma analysis was performed using a 3%/3 mm criterion with a different dose threshold. The gamma evaluation was performed within a consistent region of interest defined as the PTV for all cases. These metrics provided a quantitative measure of the deviation between the idealized IMPT dose distribution and the PSRF dose distribution, thereby characterizing the magnitude of RF crosstalk.

**Dependence of ridge-filter crosstalk on beamlet pitch**

To evaluate how beamlet lateral spacing influences the magnitude of RF crosstalk, we conducted a series of simulations using different lateral spot separations (also corresponding to the dimension of the base of each beamlet RF) of 8, 10, 12, and 15 mm and ellipsoid targets in a water phantom. The beam spot size keeps the same for each simulation. For each spacing configuration, PSRF-based dose distributions were generated and compared to the corresponding baseline IMPT plan. A 3D gamma analysis was performed to quantify discrepancies between the IMPT and PSRF dose distributions with various dose thresholds using 3%/3 mm criteria [29]. The gamma pass rate was used as a metric of agreement, with lower pass rates indicating stronger crosstalk effects.

**Iterative re-optimization for crosstalk mitigation**

As a crosstalk mitigation strategy, we propose an iterative re-optimization process (Figure 3). The optimization is driven by a loss function that quantifies the deviation of the calculated dose distribution from the clinical prescription. The loss function, which incorporates both target coverage $loss_{\text{target}}$ and organ-at-risk (OAR) sparing $loss_{\text{OAR}}$, is expressed as:

$$loss = loss_{\text{target}} + loss_{\text{OAR}},$$

where $loss_{\text{target}}$ penalizes under- or over-dosing within the planning target volume (PTV), and $loss_{\text{OAR}}$ penalizes violations of dose constraints for adjacent critical structures. These criteria were defined by the authors in our study who are practicing radiation oncologists to reflect real-world clinical standards and summarized in Table 1.

A baseline IMPT plan is first optimized by minimizing this loss function using a quasi-Newton method. Based on the optimized spot weights, an initial PSRF geometry is constructed, and the corresponding PSRF plan is calculated using MC simulation. As expected, the initial PSRF plan exhibits unwanted hot and cold spots due to RF crosstalk. To correct these hot and cold spots, we introduce a correction factor, defined as:

$$CF = D_{\mathrm{IMPT}} - 0.5 \times (D_{\mathrm{PSRF}} - D_{\mathrm{IMPT}}),$$

where $D_{\mathrm{IMPT}}$ is the reference dose distribution from the IMPT plan and $D_{\mathrm{PSRF}}$ is the calculated PSRF dose distribution. This correction factor is then incorporated into the subsequent round of optimization. The process is repeated iteratively until the dose inconsistency is reduced to a level considered clinically acceptable based on physician-defined tolerances.

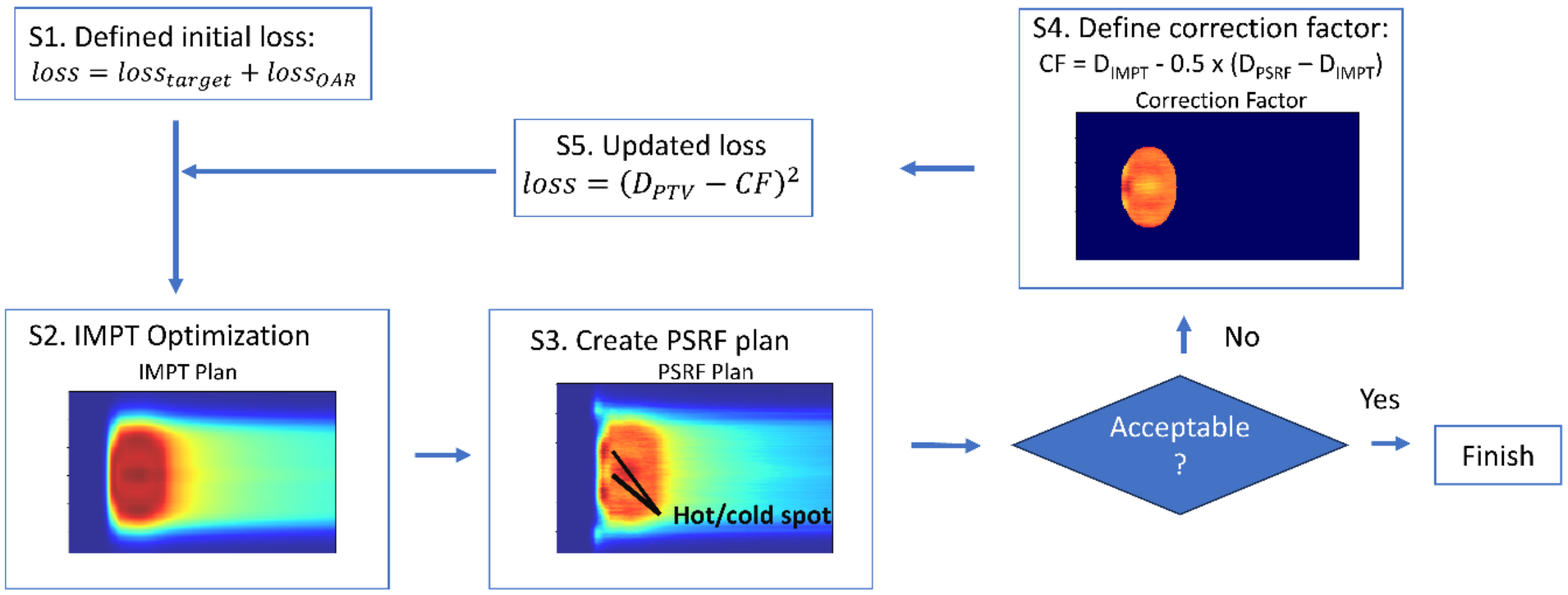

*Figure 3. Workflow of the iterative re-optimization strategy to mitigate ridge filter (RF) crosstalk. First, the initial loss function is defined as the sum of target and organ-at-risk (OAR) deviations (S1). Second, an intensity-modulated proton therapy (IMPT) plan is optimized using this loss function (S2). Third, the corresponding patient-specific ridge filter (PSRF) plan, which typically exhibits hot and cold spots due to RF crosstalk, is generated (S3). Fourth, a correction factor (CF) is defined to quantify the discrepancy between the IMPT and PSRF dose distributions (S4). Fifth, the loss function is updated accordingly (S5). This process is repeated iteratively until the dose inconsistency is reduced to an acceptable level, at which point the optimization terminates (after S3).*

**Feasibility Validation**

To demonstrate the feasibility of our method, we applied the full optimization workflow to both water phantoms and a CT scan of a pediatric osteosarcoma patient with spinal metastases. Treatment planning was performed directly on the CT geometry with clinically relevant dose constraints applied to the target and organs at risk. The baseline IMPT plan was compared with PSRF plans before and after iterative correction using DVH metrics, gamma analysis, and clinical goal evaluation, demonstrating the applicability of the proposed framework in a realistic anatomical setting.

**Beam Lateral Position Uncertainty**

To evaluate the sensitivity of the proposed delivery strategy to lateral beam mispositioning, we incorporated clinically characterized beam steering uncertainties into the dose calculation. Lateral position errors were modeled as independent Gaussian random errors applied to each scanned pencil beam spot. The error magnitude followed a normal distribution with zero mean and standard deviations of 0.26 mm in the x-direction and 0.42 mm in the y-direction, consistent with previously reported beam steering accuracy measurements[30]. For each treatment field, we generated 10 independent Monte Carlo simulations, each with a unique random lateral offset applied to all spots. The resulting perturbed dose distributions were evaluated to quantify the dosimetric impact of beam mispositioning. Across the simulation ensemble, we computed the mean and standard deviation of the dosimetric metrics of interest, providing an estimate of both the systematic and stochastic effects introduced by realistic beam steering errors.

## RESULTS

**Water phantoms**

Figure 4 shows the MC-calculated dose distributions of an optimized PSRF plan in a water phantom. Homogeneous target coverage was achieved in a cuboidal volume (6 x 6 x 4 $cm^3$) using 25 (5 x 5 grid organization) mSOBP scanning beamlets. The dose distribution was uniform and conformal in both the lateral and depth directions. For the cuboidal target configuration, where all adjacent beamlet-specific ridge filters (BSRFs) were identical, the difference between the IMPT and PSRF plans was negligible, and no crosstalk effect was observed.

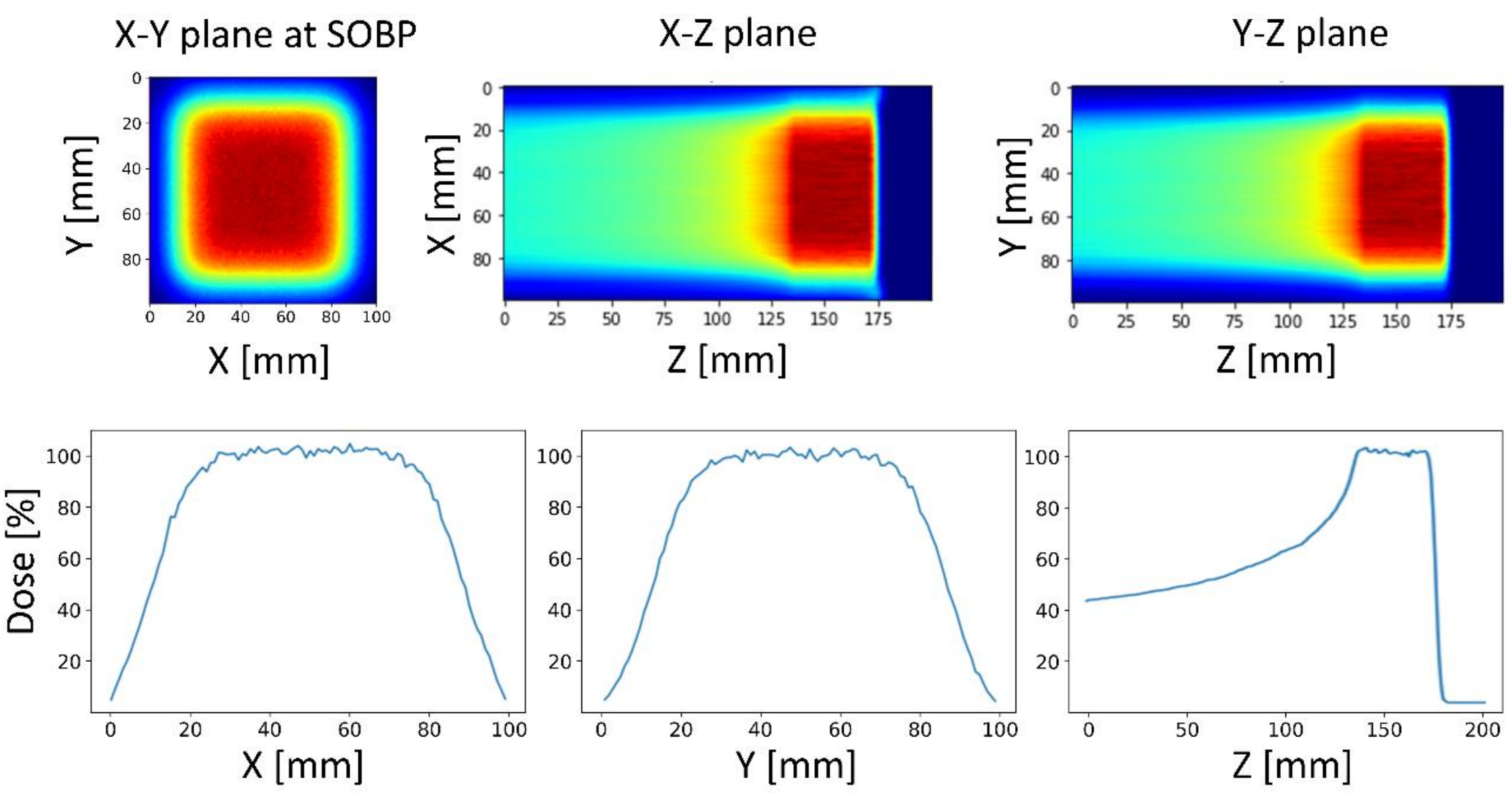

*Figure 4. Dose map and dose profiles of the patient-specific ridge filter plan in a water phantom with cuboidal target. Abbreviation: SOBP, spread-out Bragg peak*

In contrast, Figure 5 shows the results of the MC simulations for a PSRF plan in an ellipsoidal target within a water phantom. All three plans (IMPT, uncorrected PSRF, and corrected PSRF) were normalized to D95 to ensure consistent dose comparison. In this case, adjacent beamlet RFs had different shapes, and the RF crosstalk effect became substantial. The crosstalk effect in this case resulted in discrepancies between the PSRF and IMPT dose distributions, with mean relative dose differences in the target of about 8.9%. These discrepancies manifested as hot and cold spots in the dose distribution map, as well as extended tails in the dose-volume histograms (DVH) curves. For the original IMPT plan, the D95 and D5 values were 100% and 106.4% of the prescribed dose, respectively, whereas in the uncorrected PSRF plan, the D95 and D5 values were 100% and 128.2%, respectively.

Our iterative re-optimization method successfully mitigated these discrepancies into acceptable level after 3 rounds re-optimization. After correction, the value of D5 decreased to 110.8%. The mean relative dose difference between the PSRF and IMPT dose distributions within the target was reduced from 8.9% before optimization to 3.4% after optimization. Differences in the DVHs between the PSRF and IMPT plans were substantially reduced, and a gamma analysis (with a 3%/3 mm criterion) confirmed the improvement in dose conformity: before correction, gamma pass rates declined sharply at higher dose thresholds, whereas after correction, dose agreement remained consistently strong. These results demonstrate that iterative re-optimization can effectively correct RF crosstalk and restore dose conformity and accuracy.

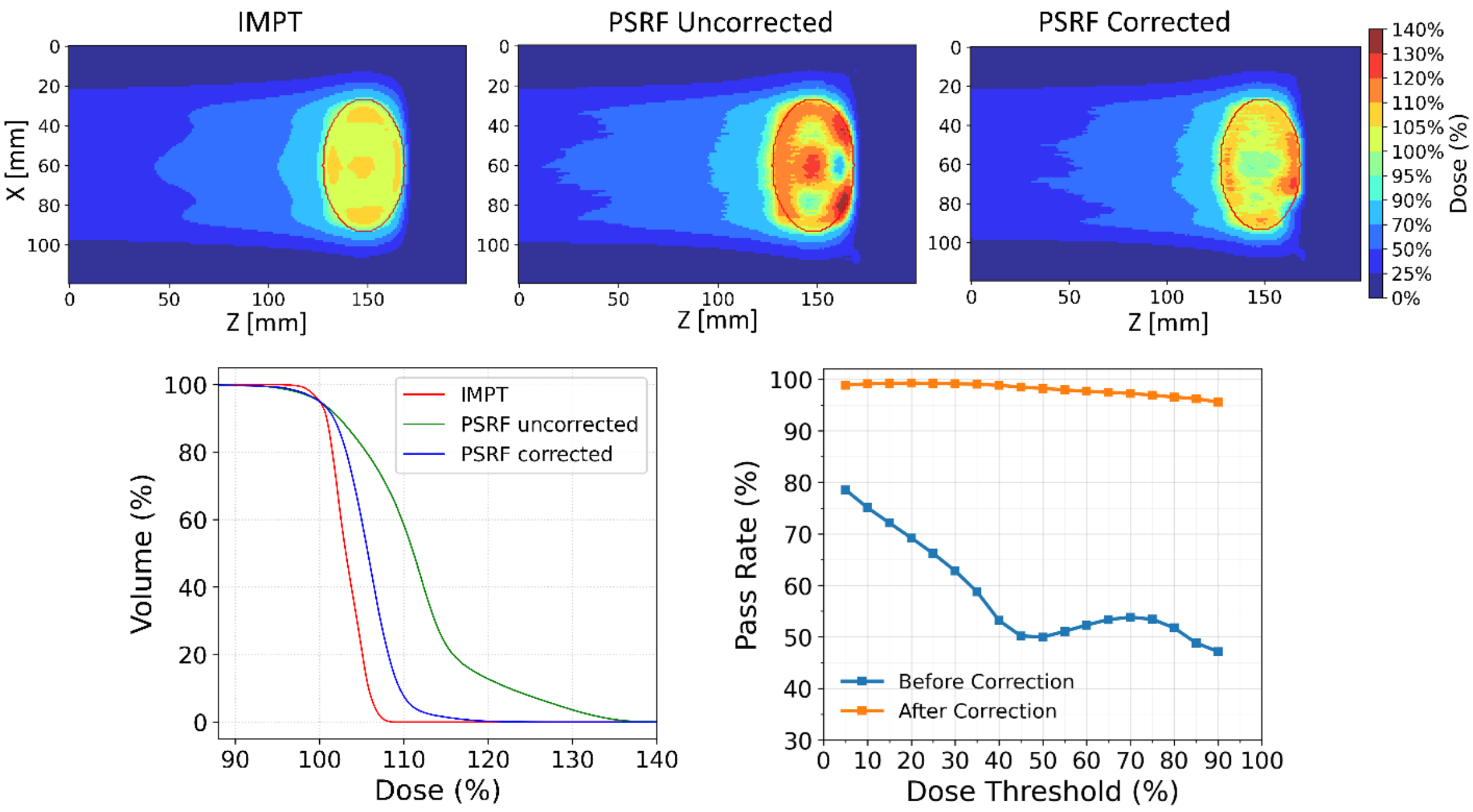

*Figure 5. Comparison of intensity-modulated proton therapy (IMPT), uncorrected patient-specific ridge filter (PSRF), and corrected PSRF plans in a water phantom with an ellipsoidal target. (Top row) Dose distributions show that the uncorrected PSRF plan, owing to ridge filter crosstalk, introduces hot and cold spots, which are mitigated after correction. (Bottom left) Dose-volume histogram curves demonstrate reduced dose accuracy in the uncorrected PSRF plan and the resulting corrected accuracy after re-optimization. (Bottom right) The gamma passing rate analysis (3%/3mm) indicates that dose discrepancies are substantially improved after correction.*

**CT geometry**

Figure 6 presents the results of a simulation within the CT geometry of a case of pediatric osteosarcoma with spinal metastases, demonstrating the clinical applicability of our proposed strategy. The optimized plan successfully met the clinical prescription criteria, confirming the feasibility of our plan in a realistic treatment context (Table 1).

| ROI/POI Name | Clinical Goal | Before Correction | After Correction |
|---|---|---|---|
| CTV | At least 80% volume at 1600 cGy | Met | Met |
| CTV | At most 2700 cGy maximum dose | Not Met | Met |
| GTV | At least 95% volume at 2100 cGy | Met | Met |
| GTV | At least 1500 cGy minimum dose | Met | Met |
| GTV | At most 2700 cGy maximum dose | Not Met | Met |
| Bag Bowel | At most 1540 cGy maximum dose | Met | Met |
| Bag Bowel | At most 5.0 $cm^3$ at 900 cGy | Met | Met |
| Kidney L | At most 67.0% volume at 800 cGy | Met | Met |
| Kidney R | At most 67.0% volume at 800 cGy | Met | Met |
| Kidneys | At most 200 $cm^3$ at 840 cGy | Met | Met |
| Liver | At most 700 $cm^3$ at 910 cGy | Met | Met |
| Skin | At most 1600 cGy maximum dose | Met | Met |
| Skin | At most 10 $cm^3$ at 1400 cGy | Met | Met |
| Spinal Cord | At most 2000 cGy maximum dose | Met | Met |

*Table 1. Summary of clinical goal evaluations before and after iterative re-optimization. Each clinical goal lists the corresponding dose constraint applied to the specified region of interest (ROI) or point of interest (POI). "Met" indicates that the dose objective was satisfied, whereas "Not Met" indicates that the dose constraint was violated.*

As illustrated in Figure 6b, the RF crosstalk effect introduced a hot spot at the distal edge of the Bragg peak (indicated by the white arrow). Following iterative re-optimization, this hot spot was effectively mitigated with 2 iterations, bringing dose levels back within the prescribed tolerances. Figure 6c compares the dose differences between the PSRF and baseline IMPT plans, showing that discrepancies were substantially reduced after correction. Figure 6d displays the DVH curves. Due to RF crosstalk, D2 within the target increased from 26.3 Gy to 30 Gy, exceeding the clinical maximum dose limit of 27 Gy. After re-optimization, D2 was reduced to 26.7 Gy, satisfying the clinical requirement. The mean relative difference within the clinical target volume (CTV) decreased from 3.7% to 1.8%, confirming a substantial improvement in dose conformity. Figure 6e shows the gamma analysis (with a 2%/2 mm criterion). The gamma analysis found that before correction, pass rates were much lower at higher dose thresholds, whereas after correction, dose agreement remained consistently high across thresholds. Together, these results demonstrate that iterative re-optimization effectively mitigates RF crosstalk and restores clinically acceptable dose distributions in patient-specific geometries.

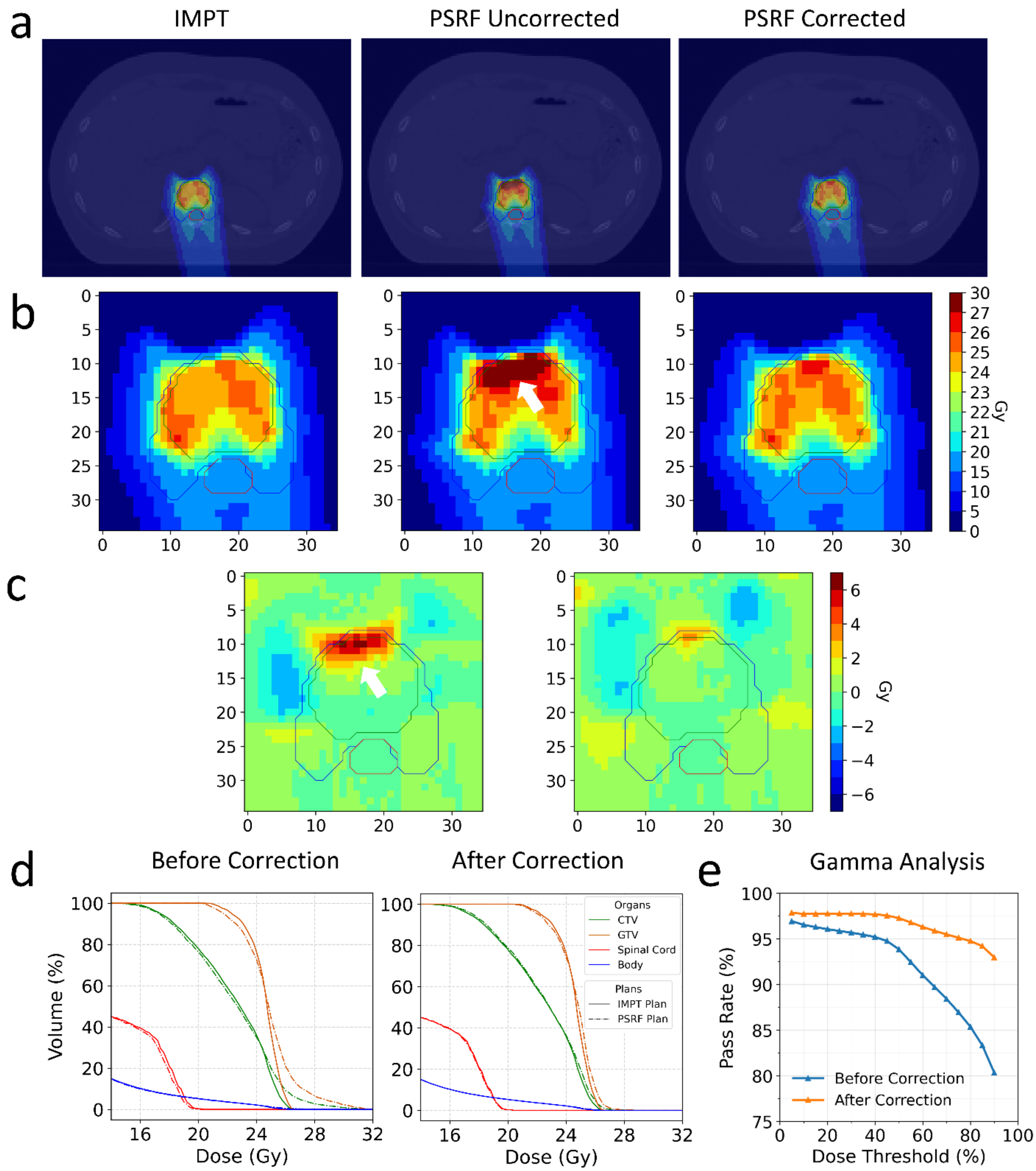

*Figure 6. Monte Carlo simulation results for a case of pediatric osteosarcoma with spinal metastases. (a) Axial dose distribution maps for intensity-modulated proton therapy (IMPT), uncorrected patient-specific ridge filter (PSRF), and corrected PSRF plans. (b) Zoomed dose distributions showing a hot spot at the distal Bragg peak due to RF crosstalk (white arrow), which was mitigated after correction. (c) Dose difference maps (i.e., showing differences between the PSRF and IMPT plans) before (left) and after correction (right). (d) Dose-volume histogram comparisons showing reduced hot spots and improved dose conformity after re-optimization (D2 reduced from 30 Gy to 26.7 Gy; the mean relative difference of dose distribution decreased from*

*3.7% to 1.8%). (e) Gamma analysis (with a 2%/2 mm criterion) showing improved agreement after re-optimization, with pass rates remaining high across dose thresholds. Abbreviations: CTV, clinical target volume; GTV, gross tumor volume.*

**Dependence of the RF crosstalk effect on lateral spacing**

To further investigate the dependence of the RF crosstalk effect on lateral beamlet spacing, we performed simulations with spacings of 8 mm, 10 mm, 12 mm, and 15 mm in water phantom with an ellipsoid target shape. All plans were normalized to D95 to ensure consistent dose comparison. The results, quantified by gamma analysis, are shown in Figure 7a. As expected, reducing the lateral spacing increased the lateral portion of beamlet passing through adjacent BSRF, thus increased the magnitude of crosstalk, as reflected by lower gamma pass rates, particularly in the high-dose regions. At smaller beamlet spacings (e.g., 8 mm), gamma pass rates are much smaller, indicating stronger interference between adjacent beamlets. Conversely, larger beamlet spacings (e.g., 15 mm) alleviated crosstalk, yielding consistently higher gamma pass rates across thresholds. Figure 7b shows the DVH curves of the PTV for different beamlet spacings. Smaller spacings exhibit longer high-dose tails, demonstrating increased RF crosstalk and reduced dose uniformity. These findings highlight the inherent trade-off between beam lateral spacing and RF crosstalk, and they emphasize the importance of novel methods, such as our iterative re-optimization approach, to resolve this limitation.

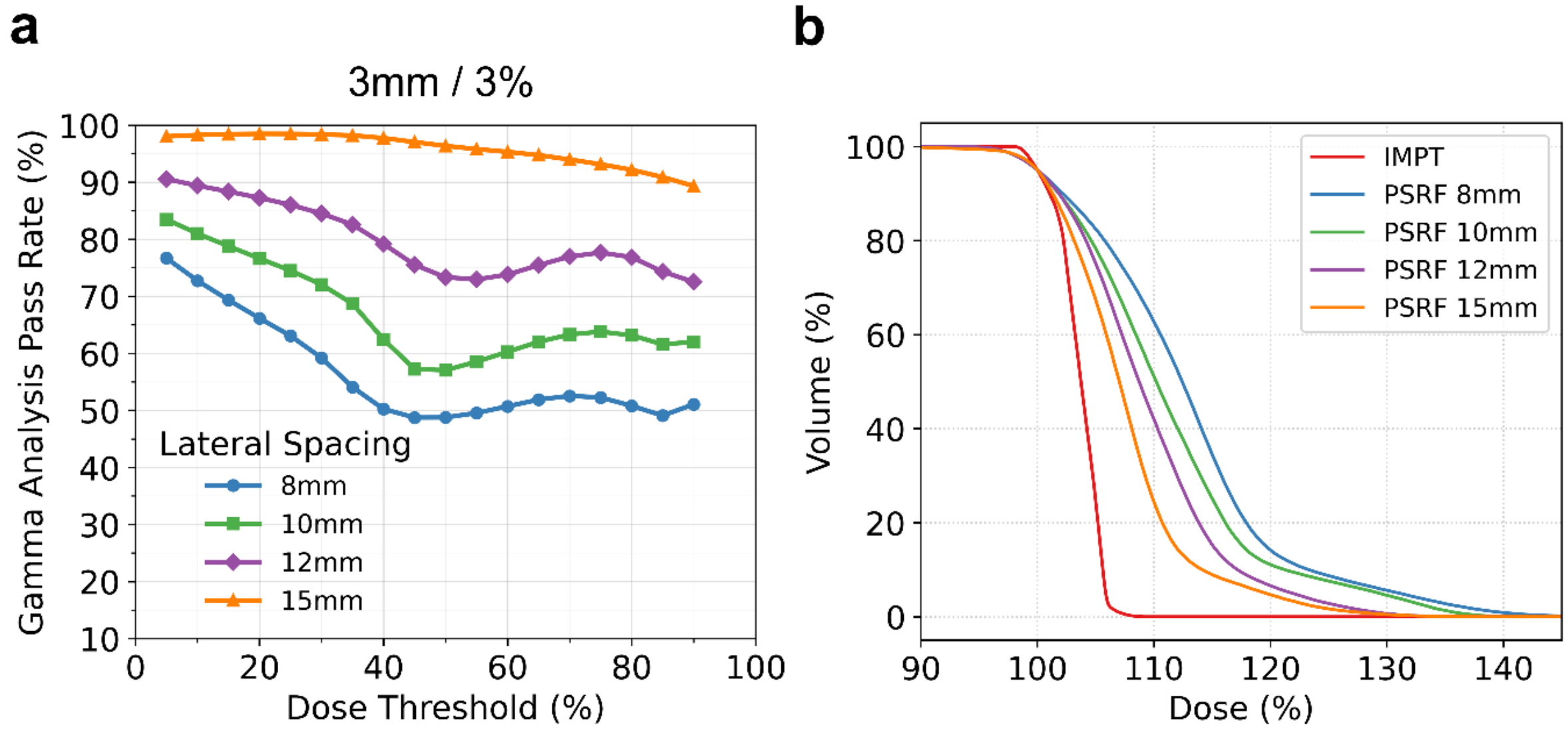

*Figure 7. (a) Gamma analysis pass rates for patient-specific ridge filter (PSRF) plans with different lateral beamlet spacings, evaluated using the 3 mm / 3 % criterion. Larger beamlet spacings resulted in higher gamma pass rates, indicating reduced RF crosstalk between neighboring beamlets. (b) Dose–volume histograms (DVHs) of the planning target volume (PTV) for different beamlet spacings. The DVH curves exhibit progressively longer high-dose tails at smaller spacings, demonstrating increased RF crosstalk and reduced dose uniformity.*

**Beam Lateral Position Uncertainty**

Figure 8 illustrates the impact of beam lateral position uncertainty on the PTV dose–volume histogram. The solid orange curve represents the nominal PSRF-corrected plan, while the solid blue curve shows the mean DVH obtained from 10 simulations with random lateral offsets applied to each pencil beam. The shaded band denotes the ±3σ envelope of these perturbed simulations, indicating the range of variation induced by realistic steering errors. Introducing lateral position errors produced only minor changes in target coverage and dose uniformity. The D95 within the PTV decreased slightly from 100% to 98.7±0.30%, and D5 increased from 110.8% to 112.7±0.50% compared with the original corrected PSRF plan. The mean relative error (MRE) between the PSRF and IMPT dose distributions within the PTV increased modestly from 3.4% to 3.9±0.002% after including beam position uncertainty, indicating that the overall agreement between the two plans remained high. Consistent with these findings, the 3D gamma analysis (3%/3 mm, 10% dose threshold) derived a passing rate of 99.8%, confirming that the dosimetric impact of the beam lateral position errors is very small and unlikely to be clinically significant.

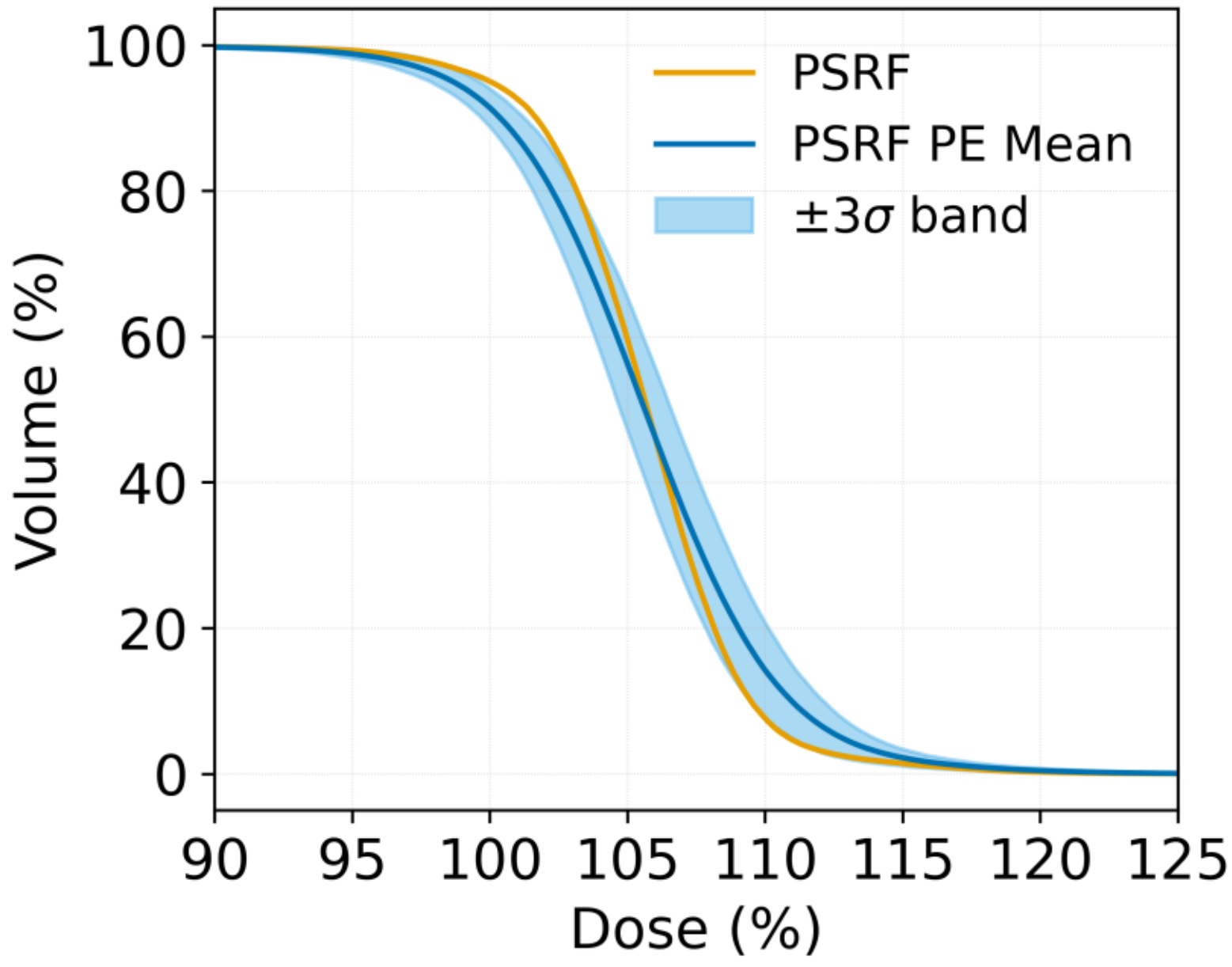

*Figure 8. Influence of beam lateral position uncertainty on the dose–volume histogram (DVH) of the PTV. The orange line shows the DVH of the nominal PSRF-corrected plan. The dark blue line represents the mean DVH from 10 Monte Carlo simulations in which random lateral position errors were applied to each pencil beam (Gaussian-distributed with $\sigma$ = 0.26 mm in x and $\sigma$ = 0.42 mm in y). The light blue shaded region indicates the ±3$\sigma$ band across these perturbed simulations, demonstrating the range of DVH variation caused by realistic beam steering inaccuracies.*

## DISCUSSION

In this study, we investigated the use of patient-specific ridge filters (PSRFs) for conformal single-energy proton dose delivery and identified ridge-filter (RF) crosstalk as a major source of dose inconsistency. While PSRF-based approaches have been proposed previously for conformal proton FLASH delivery, our results show that translating optimized spot weights from an IMPT plan into physically adjacent ridge-filter structures introduces non-negligible discrepancies that are not captured in conventional optimization frameworks. This limitation was not explicitly addressed in prior work, where emphasis has primarily been placed on demonstrating feasibility or optimizing dose and dose-rate characteristics.

The primary origin of these discrepancies is the finite lateral width of the proton beam, which leads to partial overlap of neighboring beamlets. As demonstrated in this study, the assumption that each beamlet is modulated only by its corresponding ridge-filter segment does not hold in practice.

Instead, lateral beam spread causes portions of the beam to traverse adjacent ridge structures, resulting in unintended modulation. This RF crosstalk leads to the formation of hot and cold spots in the dose distribution and degrades agreement with the baseline IMPT plan. The spatial distribution of these discrepancies and their dependence on neighboring ridge geometries support this interpretation and argue against alternative explanations such as inaccuracies in the RF definition or numerical artifacts, a concern raised by the reviewers.

One finding of this study is the strong dependence of RF crosstalk on beamlet lateral spacing. When spot size is fixed, as the spacing between beamlets decreases, the overlap between neighboring beam profiles increases, thereby amplifying inter-beamlet interaction. This behavior was consistently observed in both gamma analysis and DVH metrics, with smaller spacings producing lower gamma pass rates and more pronounced high-dose tails. Conversely, increasing the spacing reduces beam overlap and mitigates crosstalk, but at the cost of reduced beam modulation flexibility. These results demonstrate an inherent trade-off between conformality and dosimetric accuracy in PSRF-based planning, which is not captured by current conformal FLASH optimization.

This finding also helps explain why RF crosstalk has been observed in some studies but not prominently reported in others. The magnitude of the effect depends on multiple factors, including the spot size-to-beamlet spacing ratio and the complexity of the modulator geometry. In systems with smaller spot sizes or larger effective spacing, the overlap between beamlets may be limited, and crosstalk may remain negligible. In addition, when adjacent ridge filters have similar modulation patterns-such as in cases with simple geometries (e.g., cuboidal targets or targets with relatively parallel proximal and distal boundaries), the impact of RF crosstalk is further reduced because neighboring beamlets experience nearly identical modulation. In contrast, the present study operates in a regime where the beam spot size is comparable to or larger than the beamlet spacing and where the modulation varies significantly between adjacent beamlets, leading to substantial inter-beamlet overlap and observable dose discrepancies. These results suggest that RF crosstalk is not an artifact unique to this implementation but rather a general physical effect that becomes important when the beam-width-to-pitch relationship exceeds a certain threshold and when spatial variation in ridge-filter modulation is present.

To address this limitation, we developed an iterative re-optimization strategy that explicitly accounts for the discrepancy between the IMPT and PSRF dose distributions. By incorporating a correction factor into the optimization process, the method progressively compensates for crosstalk-induced deviations. The results demonstrate that this approach effectively reduces dose inconsistency, improves agreement with the IMPT reference, and restores clinically acceptable dose distributions in both phantom and patient geometries. Importantly, this framework allows the use of relatively small beamlet spacing, which is necessary for conformal coverage and flexible modulation, while mitigating the associated increase in RF crosstalk.

Despite these promising results, several limitations should be acknowledged. First, the present study is based on Monte Carlo simulations, and no experimental validation was performed, which can be necessary to confirm the magnitude of RF crosstalk and the effectiveness of the proposed mitigation strategy under realistic conditions, including potential effects of fabrication accuracy and material uncertainties. Second, only a single beam energy was considered. In practice, beam spot size and scattering characteristics vary with energy, and the relationship between beam width and spacing may differ for other energies or beamline configurations. Extending this analysis to multiple energies would improve the generalizability of the findings.

In addition, the current evaluation of robustness is limited to beam positioning uncertainty. Clinically relevant uncertainties such as patient setup errors, anatomical motion, and proton range uncertainty were not included and may further affect the delivered dose distribution. A more comprehensive robustness analysis, including these factors, would be required to fully assess clinical applicability.

Finally, although this work was motivated by conformal proton FLASH delivery, we did not explicitly evaluate dose-rate distributions. As noted by the reviewers, assessment of FLASH dose-rate metrics is important when claiming FLASH feasibility. In this study, we focused instead on the dosimetric fidelity of PSRF-based modulation, as inaccuracies in dose delivery must first be understood and controlled before meaningful evaluation of dose rate can be performed. Future work will incorporate dose-rate analysis within this framework.

**CONCLUSION**

In summary, this study identifies RF crosstalk as an important and previously under-characterized source of dose inconsistency in PSRF–based conformal proton planning. By demonstrating its dependence on beamlet spacing and providing an effective mitigation strategy, this work clarifies a key physical limitation of PSRF-based delivery and establishes a framework for improving its accuracy and robustness. These findings contribute to a better understanding of device-based modulation in proton therapy and may inform the design of more reliable conformal proton FLASH treatment approaches.